\documentclass[journal]{IEEEtran}
\usepackage[utf8]{inputenc}
\usepackage{amsmath}
\usepackage{amssymb}
\usepackage{amsthm}
\usepackage{array}
\usepackage{mathtools}
\usepackage{graphicx}
\usepackage[outdir=./]{epstopdf}
\usepackage{subcaption}
\usepackage{cleveref}
\usepackage{lipsum}
\usepackage{float}
\usepackage{tabularx}
\usepackage[bb=boondox]{mathalfa}
\usepackage[english]{babel}
\usepackage{blindtext}
\usepackage{array}
\usepackage{subfiles}
\usepackage{csquotes}
\usepackage{cite}
\usepackage{url}
\usepackage{multicol}
\usepackage{siunitx,algorithmic,bm}
\usepackage{commath}
\usepackage[section]{placeins}
\usepackage[space]{grffile}
\usepackage{blkarray}
\usepackage{breqn}
\usepackage{stackengine}
\usepackage{xcolor}

\newcommand\ubar[1]{\stackunder[1.2pt]{$#1$}{\rule{.8ex}{.075ex}}}

\newcommand{\GMat}{\ubar{\bm{G}}}
\newcommand{\GSup}{\bm{G}}

\newcommand{\FMat}{\ubar{\bm{F}}}
\newcommand{\FSup}{\bm{F}}

\newcommand{\EfMat}{\bm{E_F}}

\newcommand{\Es}{\bm{E}_s}
\newcommand{\Ee}{\bm{E}_e}
\newcommand{\Ex}{\bm{E}_{x}}

\newcommand{\WMat}{\mathbf{W}}
\newcommand{\vSig}{\mathbf{v}}

\newcommand{\AMat}{\bm{A}}
\newcommand{\BMat}{\bm{B}}

\newcommand{\nextIter}{_{j+1}}
\newcommand{\currIter}{_{j}}

\newcommand{\reals}{\mathbb{R}}
\newcommand{\naturals}{\mathbb{N}}

\newcommand{\inverse}{^{-1}}
\newcommand{\transpose}{^{T}}

\newcommand{\IMat}{\bm{I}}
\newcommand{\oneVec}{\bm{1}}
\newcommand{\Ix}{\bm{I}_x}
\newcommand{\Iu}{\bm{I}_u}

\newcommand{\eSig}{\ubar{e}}

\newcommand{\uSig}{\ubar{u}}

\newcommand{\rSig}{\ubar{r}}

\newcommand{\minimize}[2]{\underset{#1}{\text{minimize}}\quad & #2}
\newcommand{\subjectto}{\text{subject to}\quad}
\newcommand{\limit}[1]{\underset{#1\rightarrow\infty}{\text{lim}}}

\newcommand{\nx}{n_x}
\newcommand{\ns}{n_s}
\renewcommand{\ni}{n_i}
\renewcommand{\nu}{n_u}

\newcommand{\xSig}{\ubar{x}}
\newcommand{\xSup}{\mathbf{x}}
\newcommand{\zSig}{\ubar{z}}

\newcommand{\zInf}{\ubar{z}_{\infty}}
\newcommand{\zOpt}{\ubar{z}_{opt}}
\newcommand{\lambdaOpt}{\lambda_{opt}}

\newcommand{\uSup}{\mathbf{u}}

\newcommand{\xZero}{x^{0}}

\newcommand{\rSup}{\mathbf{\bar{r}}}
\newcommand{\eSup}{\mathbf{e}}

\newcommand{\dSig}{\ubar{d}}

\newcommand{\LuSup}{\bm{L}_u}
\newcommand{\LeSup}{\bm{L}_e}
\newcommand{\LxZerojSup}{\bm{L}_{x^0_j}}
\newcommand{\LxZerojPlusSup}{\bm{L}_{x^0_{j+1}}}
\newcommand{\LcSup}{\bm{L}_c}
\newcommand{\LzeroSup}{\bm{L}_0}

\newcommand{\Tuj}{\bm{T}_{u\currIter}}

\newcommand{\TxjZero}{\bm{T}_{x^0\currIter}}

\newcommand{\etaSig}{\ubar{\eta}}

\newcommand{\Az}{\bm{A}_{z}}
\newcommand{\Azj}{\bm{A_}{z\currIter}}

\newcommand{\Eu}{\bm{E}_{u}}

\newcommand{\QdeltauSup}{\bm{Q}_{\delta u}}

\newcommand{\QdeltaxSup}{\bm{Q}_{\delta x}}

\newcommand{\sx}{\ubar{\bm{s}}_{x}}
\newcommand{\sxSup}{\bm{s}_{x}}
\newcommand{\QuHat}{\bm{\hat{Q}}_{u}}
\newcommand{\QxHat}{\bm{\hat{Q}}_{x}}
\newcommand{\QeHat}{\bm{\hat{Q}}_{e}}
\newcommand{\QqHat}{\bm{\hat{\bar{Q}}}_{q}}
\newcommand{\qlHat}{\bm{\hat{\bar{q}}}_{l}}

\newcommand{\Qdeltau}{\bm{Q}_{\delta u}}
\newcommand{\Qdeltax}{\bm{Q}_{\delta x}}
\newtheorem{theorem}{Theorem}
\newtheorem{corollary}{Corollary}
\newtheorem{assumption}{Condition}
\IEEEoverridecommandlockouts

\title{Receding Horizon Iterative Learning Control for Continuously Operated Systems}

\author{Maxwell Wu$^1$, Mitchell Cobb$^{2}$, James Reed$^{3}$, Kirti Mishra$^{4}$ Chris Vermillion$^{5}$, and Kira Barton$^6$
\thanks{$^{1}$Maxwell Wu is a PhD candidate at The University of Michigan {\tt\small maxwu@umich.edu}.}%
\thanks{$^{2}$Mitchell Cobb is a is a controls engineer at Blue Origin {\tt\small mcobb@ncsu.edu}.}
\thanks{$^{3}$James Reed is an a PhD candidate at North Carolina State University {\tt\small jcreed2@ncsu.edu}.}
\thanks{$^{4}$Kirti Mishra is a Postdoctoral researcher at North Carolina State University {\tt\small kdmishra@ncsu.edu}.}
\thanks{$^{5}$Chris Vermillion is an Associate Professor in the Department of Mechanical and Aerospace Engineering at North Carolina State University, Raleigh, NC 27695, USA {\tt\small cvermil@ncsu.edu}.}
\thanks{$^{6}$Kira Barton is an Associate Professor in the Department of Mechanical Engineering at the University of Michigan, Ann Arbor, MI, 48109, USA {\tt\small bartonkl@umich.edu}.}%
}

\begin{document}
\graphicspath{{./figures/}}
\maketitle
\begin{abstract}
    
    This paper presents an iterative learning control (ILC) scheme for continuously operated repetitive systems for which no initial condition reset exists. To accomplish this, we develop a lifted system representation that accounts for the effect of the initial conditions on dynamics and projects the dynamics over multiple future iterations. Additionally, we develop an economic cost function and update law that considers the performance over multiple iterations in the future, thus allowing for the prediction horizon to be larger than just the next iteration. Convergence of the iteration varying initial condition and applied input are proven and demonstrated using a simulated servo-positioning system test case.

\end{abstract}

\section{Introduction}
\begin{figure}[ht]
    \centering
    \includegraphics[width=\columnwidth]{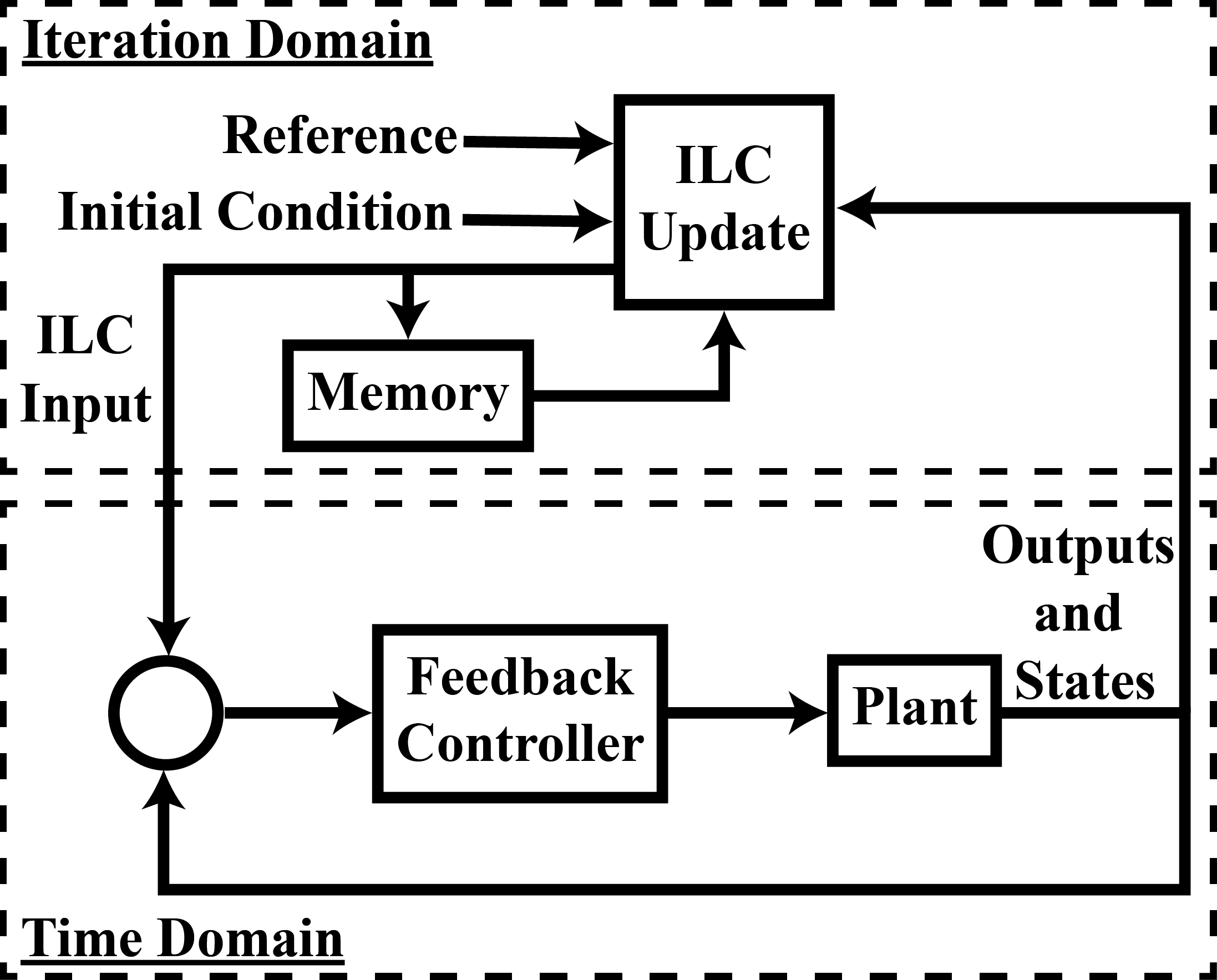}
    \caption{ILC utilizes iteration domain feedback to inform how a system is to be controlled in future iterations. An iteration-varying feedforward input signal aims to improve the behavior of the system by compensating for model uncertainties.}
    \label{fig:RHILC_Block_Diagram}
\end{figure}

\IEEEPARstart{F}{or} systems that exhibit repetitive behavior, repetitive and iterative learning control (ILC) techniques have proven to be useful tools that enable system performance to be improved through a learning-based update of the control signal. Traditionally, these strategies have aimed to eliminate tracking error of a known reference signal by leveraging information available in previous executions of a task to counteract uncertainties in the system. Typically, ILC has been developed for batch processes wherein the system undergoes an initial condition reset between iterations of the repetitive task. A common method of implementation of ILC controllers is depicted in the block diagram shown in Figure \ref{fig:RHILC_Block_Diagram}. Here, measurements of the system states and outputs, as well as knowledge of previously applied control signals, initial conditions, and reference signals, are utilized to design improved control signals in future iterations \cite{Bristow2006}. Meanwhile, repetitive control has generally been used for continuous processes where no such reset of the initial condition occurs between iterations \cite{Wang2009}.

Many repetitive systems operate continuously. For instance, in autonomous racing applications the vehicle must repeatedly follow a predefined path with the goal of minimizing total lap time. Here, the control actions at a given lap will influence the behavior of the vehicle in future laps. Specifically, the system is considered continuously operable in that the initial condition at each iteration is given by the terminal condition at the previous iteration. Additionally, note that here the performance is not strictly dependent on the ability of the system to accurately track a predefined reference trajectory. Rather, the improved performance is achieved if the system is better able to minimize the non-traditional performance metric of lap time. These \textit{economic} metrics enable system performance to be assessed for a broader range of systems such as robotic prosthetic legs that aim to follow a continuous gait trajectory while expending as little energy as possible, or tethered energy systems that follow a repetitive closed flight path with the primary objective of maximizing power generation.

Traditionally, repetitive control design has relied on a frequency domain analysis wherein the internal model principle is employed for the purpose of tracking periodic references or rejection of periodic disturbances \cite{Wang2009}. However, while suitable for strict trajectory tracking objectives, controller design in the frequency domain is difficult to exercise when improvement in more general economic performance metrics is desired. On the other hand, ILC design has often leveraged state space models to improve system performance through iteration-based feedback. The use of state-space analysis has enabled the development of system representations in the lifted domain, as well as `norm-optimal' controller design, which permits the intuitive construction of quadratic cost functions as a function of the control signal \cite{Gunnarsson2001}.

Historically, ILC designs have been made to minimize these cost functions by updating the feedforward control signal from data obtained in the previous iteration. While such strategies have proven useful, by only using information from a single iteration, these techniques neglect potential benefits to system performance that can be obtained by utilizing information from multiple iterations. To combat this, a subfield of ILC called `predictive iterative learning control', as described in \cite{chu2015iterative}, \cite{wang2011predictive}, \cite{amann1998predictive},
\cite{arif2000prediction}, and 
\cite{chu2015predictive}, has been developed that not only uses information from the previous iteration, but also uses predictions of system behavior in future iterations to update the control signal. These strategies have shown improved convergence rates in comparison to traditional ILC while still establishing requirements for stability. However, these works consider a class of systems where the initial condition is reset between iterations.

To accommodate continuously operable systems, several techniques have been created. The strategy given by \cite{borelli1} and \cite{borelli2} iteratively updates safe sets over which system constraints are satisfied. A model predictive control optimization problem is then solved where the enlarged safe sets allow for a broader selection of control signals to be applied, thus giving opportunities for improved performance. This strategy has been previously used for the autonomous racing application described above and shown to be effective in reducing total lap time. However, this formulation dispenses with the `lifted system' representation commonly used in ILC, as well as learning filters used to construct a closed form update law. In \cite{Lee2001} and \cite{Gupta2006}, a state-space based repetitive control strategy is designed to minimize performance of a system over an infinite prediction horizon by describing the problem as an iteration-domain LQ regulation problem with a control law given by solution of the algebraic Riccati equation. However, only the infinite horizon case is considered and no analysis giving requirements for stability or robustness to disturbances is given.

To address these issues, a receding horizon ILC formulation is developed for application to continuously operated systems. The contributions of this paper are given by
\begin{enumerate}
    \item Generation of a lifted system representation that describes the impact of initial conditions and a multi-iteration control signal on the system dynamics.
    \item A closed form update law that describes how the control signal is updated to reduce the value of a multi-iteration cost function.
    \item An analysis of conditions for closed loop stability and desired system performance with considerations toward robustness to uncertain plant dynamics and disturbances.
    \item Implementation of the algorithm to linear time-invariant and linear time varying systems.
\end{enumerate}

\section{Receding Horizon Iterative Learning Control} \label{sec:RHILC}

The proposed ILC formulation contains two features that facilitate the consideration of continuous operation where an initial condition reset between cycles does not exist. This is achieved by combining a lifted system model that explicitly includes the impact of nonzero initial conditions with a performance index that includes predicted performance over multiple future iterations. Thus, the framework is capable of implicitly considering the impact of control decisions from one iteration on future iterations.  The resulting iterative update law then provides control sequences for multiple future iterations.  However, due to the need to adjust for disturbances and modeling uncertainties and inaccuracies, this feedforward control sequence is only applied one iteration at a time before being updated. The following subsections detail this approach.

\subsection{Multi-Iteration System Model}\label{sec:MultiIterationSystemModel}
We first consider systems with discrete time dynamic models of the form
\begin{align}
    x^{k+1}\currIter = \AMat x^k\currIter+\BMat u^k\currIter
    \label{Eq:LTI State Space Model}
\end{align}
with state vector $x\in\reals^{\nx}$ and control input $u\in\reals^{\nu}$ where $k\in\naturals$ is a timestep index, $j\in\naturals$ is an iteration index, and $\nx$ and $\nu$ indicate the number of states and inputs. We specifically consider the case where $\nu=1$. $\AMat$ and $\BMat$ are appropriately sized, real valued matrices. We now define the lifted vectors
\begin{align}
\label{Eq: Lifted Vector Definitions}
\begin{split}
    \xSig\currIter\triangleq\begin{bmatrix}(x^1\currIter)\transpose &  (x^2\currIter)\transpose & \hdots &  (x^{\ns-1}\currIter)\transpose &  (x^{\ns}\currIter)\transpose\end{bmatrix}\transpose\\
    \uSig\currIter\triangleq\begin{bmatrix}(u^0\currIter)\transpose &  (u^1\currIter)\transpose & \hdots &  (u^{\ns-2}\currIter)\transpose &  (u^{\ns-1}\currIter)\transpose\end{bmatrix}\transpose,
\end{split}
\end{align}
which give the state and input sequences over iteration $j$. Using this notation gives the lifted system model
\begin{align}
    \xSig\currIter=\GMat\uSig\currIter+\FMat\xZero\currIter
    \label{Eq: Lifted Dynamics xZero}
\end{align}
where $\xZero\currIter$ denotes the initial condition at iteration $j$ and $\GMat\in\reals^{\ns\nx\times\ns}$ is a block matrix where the block element $\ubar{G}_{m,p}\in\reals^{\nx}$ in the $m^{th}$ block row and $p^{th}$ column of $\GMat$ is given by
\begin{equation}
\label{Eq: Gmp Definition}
    \ubar{G}_{m,p} = 
    \begin{cases}
    \mathbb{0}^{n_x}              & m < p         \\

    \AMat^{m-p}\BMat    & m\geq p.
    \end{cases}
\end{equation}
Similarly, the block element in the $m^{th}$ block row of $\FMat\in\reals^{\ns\nx\times\nx}$ is given by
\begin{equation}
\label{Eq: Fm Definition}
    \ubar{F}_{m} = \AMat^m.
\end{equation}
As the system is considered to be operated continuously, the initial conditions are not reset between iterations. More specifically, the terminal condition from iteration $j$ is equal to the initial condition at iteration $j+1$. Consequently, with a shift in the iteration index, (\ref{Eq: Lifted Dynamics xZero}) can be alternatively expressed as
\begin{align}
    \xSig\nextIter=\GMat\uSig\nextIter+\FMat\EfMat\xSig\currIter
    \label{Eq: Lifted Dynamics xj}
\end{align}
where 
\begin{align}
    E_F &\triangleq \begin{bmatrix} \mathbb{0}^{n_x \times n_x(n_s-1)} & \mathbb{I}^{n_x \times n_x} \end{bmatrix},
\end{align}
which is used to select the terminal states from the $\xSig\currIter$ vector.

For causal systems that lack an initial condition reset, the control sequence chosen for one iteration has a direct impact on the system dynamics at subsequent iterations. As a result, control decisions made at the current iteration will have effects on system performance in future iterations. To capture these effects, a \textit{multi-iteration} dynamic model is developed to estimate the behavior of the system over a prediction horizon. This estimate of the system dynamics is then used to derive learning filters.

Before describing the multi-iteration dynamic model, a standard nomenclature is introduced. First, note that $\xSig_j$ and $\uSig_j$ are referred to as ``lifted'' vectors. These vectors contain data over a \textit{single} iteration and are formed by concatenating a sequence of system states or controls in chronological order as in \eqref{Eq: Lifted Vector Definitions}. Additionally, the matrices $\GMat$ and $\FMat$ are referred to as ``lifted'' system matrices in that they describe the system's behavior over one iteration in response to $\uSig_j$ and $\xZero\currIter$. However, to enable the characterization of the system dynamics over multiple iterations, this lifted notation in insufficient. Instead, to generate a multi-iteration dynamic model, it will be necessary to concatenate multiple instances of the lifted vectors and matrices. These further concatenated vectors and matrices are then referred to as ``super-lifted'' to emphasize that they are constructed from the single-iteration lifted vectors and matrices. Suppose we are attempting to capture the system behavior over $\ni$ future iterations. The super-lifted state and control vectors are then defined as
\begin{align}
\xSup\nextIter \triangleq \begin{bmatrix} \xSig\nextIter \\ \xSig_{j+2} \\ \vdots \\ \xSig_{j+\ni-1} \\ \xSig_{j+\ni}\end{bmatrix}, \quad \mathrm{and} \quad
\uSup\nextIter \triangleq \begin{bmatrix} \uSig\nextIter \\ \uSig_{j+2} \\ \vdots \\ \uSig_{j+\ni-1} \\ \uSig_{j+\ni}\end{bmatrix},
\end{align}
respectively. The relationship between $\xSup\nextIter$ and $\uSup\nextIter$ is then given as
\begin{align}
    \label{Eq: Super Lifted Model}
    \xSup\nextIter=\GSup\uSup\nextIter+\FSup\xZero\nextIter
\end{align}
where
\begin{align}
    \GSup\triangleq
    \begin{bmatrix}
    \GMat & \bm{0} & \hdots & \hdots & \bm{0}\\
    \FMat\EfMat\GMat & \GMat & \bm{0} & \hdots & \bm{0}\\
    \vdots & \vdots & \vdots & \ddots & \vdots\\
    (\FMat\EfMat)^{\ni-2}\GMat & (\FMat\EfMat)^{\ni-3} & \hdots & \GMat & \bm{0}\\
    (\FMat\EfMat)^{\ni-1}\GMat & (\FMat\EfMat)^{\ni-2} & \hdots & \FMat\EfMat\GMat & \GMat
    \end{bmatrix}
\end{align}
and
\begin{align}
    \FSup\triangleq
    \begin{bmatrix}
    \FMat \\
    \FMat\EfMat\FMat\\
    \vdots\\
    (\FMat\EfMat)^{\ni-2}\FMat\\
    (\FMat\EfMat)^{\ni-1}\FMat
    \end{bmatrix}.
\end{align}
Additionally, we define the super-lifted error vector as
\begin{align}
\label{Eq: Super Error Definition}
    \eSup\nextIter \triangleq \Ix\rSig-\xSup\nextIter
\end{align}
where $\rSig$ is a lifted reference signal vector constructed from an ordered set of reference states, $\{r_1,...,r_{\ns-1},r_{ns}\}$ where $r_q\in\reals^{\nx}$ according to the same method as shown in \eqref{Eq: Lifted Vector Definitions}. Additionally, $\IMat_{(\cdot)}$ is defined as the vertical concatenation of $\ni$ copies of an appropriately sized identity matrix $\mathbb{I}$. 


A performance index, $\mathbf{J}_{j+\ni} $, that accounts for system behavior over $n_i$ iterations is then generated as
\begin{equation}\label{eqn:longFormPerformance}
\begin{split}
\mathbf{J}_{j+\ni} 
\!\!=\!\! \sum_{k=1}^{\ni}(
& \uSig_{j\!+\!k}^T \!Q_u\! \uSig_{j\!+\!k} \!\!+\!\! \left(\uSig_{j\!+\!k}\!\!-\!\!\uSig_{j\!+\!k\!-\!1}\right)^T \!\!Q_{\delta u}\! \left(\uSig_{j\!+\!k}\!\!-\!\!\uSig_{j\!+\!k\!-\!1}\right) \\
& +\!\! \xSig_{j\!+\!k}^T \!Q_x\! \xSig_{j\!+\!k} \!\!+\!\! \left(\xSig_{j\!+\!k}\!\!-\!\!\xSig_{j\!+\!k\!-\!1}\right)^T \!\!Q_{\delta x}\! \left(\xSig_{j\!+\!k}\!\!-\!\!\xSig_{j\!+\!k\!-\!1}\right) \\
& +\!\!\underline{e}_{j\!+\!k}^T \!Q_e\! \underline{e}_{j\!+\!k} \!\!+\!\! 2\sx \xSig_{j\!+\!k})
\end{split}
\end{equation}
where $\eSig_{j+k}\triangleq\rSig-\xSig_{j+k}$. Additionally, $Q_u$, $Q_{\delta u}$, $Q_x$, $Q_{\delta x}$, and $Q_e$ are diagonal, positive, semi-definite weighting matrices applied to the quadratic terms and are defined in Appendix \ref{apn:perfIndexWeights}. 

The last term of \eqref{eqn:longFormPerformance} is included to accommodate more general economic objectives. Here, $\sx\in \mathbb{R}^{n_x n_s}$ represents a weighted linear approximation of a user-defined economic performance metric, $J_e(\xSig,\uSig)$. Specifically, given a nominal sequence of states, $\{x_l(0),...x_l(n_s-1)\}$, and inputs, $\{u_l(0),...,u_l(n_s-1)\}$, $\sx$ is defined as 
\begin{equation}
    \sx \triangleq  \begin{bmatrix} \nabla_{x} J_e(x,u)\rvert_{\begin{subarray}{l}x=x_l(0) \\ u=u_l(0)\end{subarray}} \\
    \vdots\\
    \nabla_{x} J_e(x,u)\rvert_{\begin{subarray}{l}x=x_l(n_s-1) \\ u=u_l(n_s-1)\end{subarray}}
    \end{bmatrix}^T Q_{sx},
\end{equation}
where $Q_{sx}$ is a diagonal weighting matrix designed to weight and also normalize the states in the overall performance.

By defining the weighting matrices $\mathbf{Q_u}$, $\QdeltauSup$, $\mathbf{Q_x}$, $\QdeltaxSup$, $\mathbf{Q_e}$ as in Appendix \ref{apn:perfIndexWeights} and $\sxSup\triangleq\Ix\sx$, the performance index given by \eqref{eqn:longFormPerformance} can be re-expressed in an equivalent super-block structure as a function of the super-lifted vectors $\uSup_{j+1}$, $\xSup_{j+1}$, and $\eSup\nextIter$ according to
\begin{equation}\label{eqn:superBlockPerformance}
\begin{split}
\mathbf{J}_{j+\ni} = 
& \mathbf{u}_{j+1}^T \mathbf{Q_u} \mathbf{u}_{j+1} + \mathbf{u}_{j+1}^T \mathbf{D}_u^T \QdeltauSup \mathbf{D}_u \mathbf{u}_{j+1} \\
&+ \left(\mathbf{E}_u\mathbf{u}_{j+1}-\uSig\currIter\right)^T Q_{\delta u} \left(\mathbf{E}_u\mathbf{u}_{j+1}-\uSig\currIter\right)\\
&+ \mathbf{x}_{j+1}^T \mathbf{Q_x} \mathbf{x}_{j+1} + \mathbf{x}_{j+1}^T \mathbf{D}_x^T \QdeltaxSup \mathbf{D}_x \mathbf{x}_{j+1} \\
&+ \left(\mathbf{E}_x\mathbf{x}_{j+1}-\xSig\currIter\right)^T Q_{\delta x} \left(\mathbf{E}_x\mathbf{x}_{j+1}-\xSig\currIter\right)\\
&+ \mathbf{e}_{j+1}^T \mathbf{Q_e} \mathbf{e}_{j+1}+2\sxSup\mathbf{x}_{j+1}.
\end{split}
\end{equation}
Here, $\mathbf{D}_u\in \mathbb{R}^{(\ni-1) n_s\times \ni n_s}$ and $\mathbf{D}_x\in \mathbb{R}^{(\ni-1) n_s \nx\times \ni n_s \nx}$ are difference operator matrices that calculate the difference between input and state sequences in subsequent iterations according to 
\begin{equation}
    \mathbf{D}_{(\cdot)} = \begin{bmatrix}
    \mathbb{I} & -\mathbb{I} & \mathbb{0} & \hdots\\
    \mathbb{0} & \mathbb{I} & -\mathbb{I} &  \hdots\\
    \vdots & \vdots & \vdots & \ddots
    \end{bmatrix},
\end{equation}
with $\mathbb{I}\in \reals^{n_{(\cdot)} n_s \times n_{(\cdot)} n_s}$.  Additionally, the matrices $\mathbf{E}_u$ and $\mathbf{E}_x$, are operation matrices that select the first lifted vector from the super-vector.  Specifically, 
\begin{equation}
    \mathbf{E}_{(\cdot)}\triangleq \begin{bmatrix} \mathbb{I}^{n_{(\cdot)} n_s \times n_{(\cdot)} n_s} & \mathbb{0}^{n_{(\cdot)} n_s \times (N_i-1) n_{(\cdot)} n_s} \end{bmatrix}.
\end{equation}
Substituting \eqref{Eq: Super Lifted Model} and \eqref{Eq: Super Error Definition} into \eqref{eqn:superBlockPerformance}, allows for $\mathbf{J}_{j+\ni}$ to be expressed solely as a function of the known quantities $\uSig\currIter$, $\xZero\currIter$, and $\xZero\nextIter$, as well as the design parameter $\uSup\nextIter$. 

We can obtain an expression for the performance index completely in terms of the known quantities from the last iteration, $\uSig_j$, and $\xSig_j$, along with the control sequences over the next several iterations, $\mathbf{u}_{j+1}$. Similar to the strategy employed in \cite{p2p4}, by taking the gradient of $\mathbf{J}_{j+1}$ with respect to $\mathbf{u}_{j+1}$, setting the result equal to the zero vector, and re-arranging the resulting expression, we obtain the update law 
\begin{align}
    \uSup\nextIter=\LuSup\uSig\currIter+\LeSup\eSig\currIter+\LxZerojSup\xZero\currIter+\LxZerojPlusSup\xZero\nextIter+\LcSup
    \label{Eq: uSup Update Law}
\end{align}
where
\begin{align}
    \begin{split}
        \LzeroSup \triangleq&\QuHat+\GSup\transpose(\QxHat+\QeHat)\GSup\\
        \LuSup\triangleq&\LzeroSup\inverse(\Eu\transpose\Qdeltau+\GSup\transpose\QeHat\Ix\GMat+(\Ex\GSup)\transpose\Qdeltax\GMat)\\
        \LeSup\triangleq&\LzeroSup\inverse\GSup\transpose\QeHat\Ix\\
        \LxZerojSup\triangleq&\LzeroSup\inverse((\Ex\GSup)\transpose\Qdeltax\FMat+\GSup\QeHat\Ix\FMat)\\
        \LxZerojPlusSup\triangleq&-\LzeroSup\inverse\GSup\transpose(\QxHat+\QeHat)\FSup\\
        \LcSup\triangleq&-\LzeroSup\inverse\GSup\sxSup
    \end{split}
    \label{Eq: Learning Filter Definitions}
\end{align}
and
\begin{align}
    \mathbf{\hat{Q}_u} & \triangleq \mathbf{Q_u} + \mathbf{D}_u^T\QdeltauSup \mathbf{D}_u + \mathbf{E}_u^T Q_{\delta u} \mathbf{E}_u\\
    \mathbf{\hat{Q}_x} & \triangleq \mathbf{Q_x} + \mathbf{D}_x^T\QdeltaxSup \mathbf{D}_x + \mathbf{E}_x^T Q_{\delta x} \mathbf{E}_x \\
    \mathbf{\hat{Q}_e} & \triangleq \mathbf{Q_e}.
\end{align}
Hence, as depicted in the block diagram in Figure \ref{fig:RHILC_Block_Diagram}, the ILC update utilizes information from the current input sequence, as well as the error signal and initial conditions to generate the feedforward input sequence at the next iteration.

Note that the update law given by \eqref{Eq: uSup Update Law} generates the super-lifted vector $\uSup\nextIter$ that describes the optimal control sequence to be applied over the next $n_i$ iterations. While one approach might be to apply the control sequence given by $\uSup\nextIter$ over the next $n_i$ iterations, an iteration-domain receding horizon approach is instead proposed where only the control sequence corresponding to the next iteration is applied before recomputing the optimal super-lifted control sequence. In other words, the control sequence given by
\begin{align}
    \label{Eq: uSig from uSup}
    \uSig\nextIter=\Eu\uSup\nextIter
\end{align}
is applied over iteration $j+1$. This strategy mimics the control law used in generalized predictive control as described in \cite{Clarke1987}, as well as model predictive control. By updating the control sequence after every iteration, we incorporate the most recent information, making the system more robust to disturbances and modeling uncertainties.  We describe the resulting control structure as iteration-domain receding horizon control or receding horizon iterative learning control (RHILC).


\section{Controller Analysis} \label{sec:ControllerAnalysis}

The stability properties and desired system behavior of the RHILC controller are now examined.

\subsection{Stability Analysis} \label{sec:StabilityAnalysis}

We consider the application of the RHILC algorithm to a plant with dynamics given as
\begin{align}
    \label{Eq: Plant Dynamics}
    x^{k+1}\currIter=\AMat\currIter x\currIter^k+\BMat\currIter u\currIter^k+d^k\currIter.
\end{align}
where $\AMat\currIter$ and $\BMat\currIter$ denote the iteration varying plant dynamics and $d\currIter\in\reals^{\nx}$ is a disturbance. From \eqref{Eq: Gmp Definition} and \eqref{Eq: Fm Definition}, the plant dynamics can be expressed in lifted form as
\begin{align}
    \label{Eq: Lifted Plant Dynamics}
    \xSig\currIter=\GMat\currIter\uSig\currIter+\FMat\currIter\xZero\currIter+\dSig\currIter
\end{align}
where $\GMat\currIter$ and $\FMat\currIter$ are iteration varying plant matrices and $\dSig\currIter\in\reals^{\ns\nx}$ captures the effect of the disturbance on the state sequence.

Since the system operates continuously such that $\xZero\nextIter=\EfMat\xSig\currIter$, substitution of \eqref{Eq: Lifted Plant Dynamics} yields that the initial condition at each iteration evolves according to
\begin{align}
    \xZero\nextIter=\EfMat\bigg(\GMat\currIter\uSig\currIter+\FMat\currIter\xZero\currIter+\dSig\currIter\bigg).
    \label{Eq: xZero Dynamics}
\end{align}

Consequently, substituting \eqref{Eq: Lifted Plant Dynamics} and \eqref{Eq: xZero Dynamics} into \eqref{Eq: uSig from uSup} yields
\begin{align}
\label{Eq: uSig Dynamics}
\begin{split}
    \uSig&\nextIter=\\
    &\Tuj\uSig\currIter+\TxjZero\xZero\currIter+\Eu\bigg(\LeSup\rSig+(\LxZerojPlusSup\EfMat-\LeSup)\dSig\currIter+\LcSup\bigg)
\end{split}
\end{align}
with 
\begin{align}
    \Tuj&=\Eu\bigg(\LuSup+\LxZerojPlusSup\EfMat\GMat\currIter-\LeSup\GMat\currIter\bigg)\\
    \TxjZero&=\Eu\bigg(\LxZerojPlusSup\EfMat\FMat\currIter+\LxZerojSup-\LeSup\FMat\currIter\bigg).
\end{align}

We then define the vector $\zSig\currIter$ as
\begin{align}
    \zSig\currIter\triangleq\begin{bmatrix}
    \uSig\currIter\\ \xZero\currIter
    \end{bmatrix},
\end{align}
which is the concatenation of the lifted input vector and initial condition at iteration $j$.

From \eqref{Eq: uSig Dynamics} and \eqref{Eq: xZero Dynamics}, this then gives
\begin{align}
    \zSig\nextIter=\Azj\zSig\currIter+\etaSig\currIter
    \label{Eq: zSig Dynamics}
\end{align}
where
\begin{align}
    \Azj&\triangleq\begin{bmatrix}
    \Tuj & \TxjZero\\
    \EfMat\GMat\currIter & \EfMat\FMat\currIter
    \end{bmatrix}\\
    \etaSig\currIter&\triangleq
    \begin{bmatrix}
    \Eu(\LeSup\rSig+(\LxZerojPlusSup\EfMat-\LeSup)\dSig\currIter+\LcSup)\\ \EfMat\dSig\currIter
    \end{bmatrix}.
\end{align}
Hence, the evolution of the input sequence and initial condition can be described as a linear system.

The requirements for robust stability of the closed-loop system are now presented. Here, let $\rho(\mathbf{\Lambda})$ denote the spectral radius of square matrix $\Lambda$.
\begin{assumption}
\label{Cond: Az Spectral Radius}
$\rho(\Azj)<1$ for all $j$.
\end{assumption}
\begin{assumption}
\label{Cond: Convergence} $\limit{j}\Azj=\Az$ and $\limit{j}\dSig\currIter=\dSig$ for constants $\Azj$ and $\dSig$.
\end{assumption}
Condition \ref{Cond: Az Spectral Radius} is a general requirement for stability of linear systems. Condition \ref{Cond: Convergence} states that the dynamics of the plant, as well as the disturbance signal, converge over the iteration domain.
\begin{corollary}
\label{Corr: Exponential Stability}
If Condition \ref{Cond: Az Spectral Radius} holds, the system given by \eqref{Eq: zSig Dynamics} is exponentially stable.
\end{corollary}
Corollary \ref{Corr: Exponential Stability} is obtained from standard linear systems theory. Additionally, the converged value of $\zSig\currIter$ can be assessed.
\begin{theorem}
If Conditions \ref{Cond: Az Spectral Radius} and \ref{Cond: Convergence} hold, then
\begin{align}
    \limit{j}\zSig\currIter=(\IMat-\Az)\inverse\etaSig \text{ where } \etaSig=\etaSig\currIter(\dSig)
\end{align}
\end{theorem}
\begin{proof}
Taking the limit of (\ref{Eq: zSig Dynamics}) gives
\begin{align}
    \limit{j}\zSig\nextIter=\limit{j}\left(\Azj\zSig\currIter+\etaSig\currIter\right)
\end{align}
Applying the assumptions on $\Azj$ and $\dSig\currIter$ gives
\begin{align}
    \limit{j}\zSig\nextIter=\limit{j}\zSig\currIter=\limit{j}\left(\Az\zSig\currIter+\etaSig\right)
\end{align}
where
\begin{align*}
    \etaSig\triangleq\etaSig\currIter(\dSig),
\end{align*}
which implies that
\begin{align}
    \limit{j}(\IMat-\Az)\zSig\currIter=\etaSig.
\end{align}
Thus,
\begin{align}
    \zSig_\infty\triangleq\limit{j}\zSig\currIter=(\IMat-\Az)\inverse\etaSig
    \label{Eq: Definition of zinf}
\end{align}
as a consequence of $\rho(\Az)<1$.
\end{proof}
Hence, under satisfaction of Conditions \ref{Cond: Az Spectral Radius} and \ref{Cond: Convergence}, a closed form for the converged value of the input sequence and initial condition can be generated.
\subsection{Desired Converged Controller Behavior} \label{sec:DesiredBehavior}
From Section \ref{sec:StabilityAnalysis}, requirements for robust convergence of the input sequence and initial condition are established. While meeting Conditions \ref{Cond: Az Spectral Radius} and \ref{Cond: Convergence} establish sufficient criterion for convergence, note that the converged value of the input sequence and initial condition as given by $\zInf$ is not necessarily the minimizer of $\mathbf{J}_{j+\ni}$. Hence, we now investigate what the input sequence and initial condition should be at convergence to achieve desired system behavior. For this, the operation matrices $\Es\in\reals^{\ns\nu(\ns\nu+\nx)}$ and $\Ee\in\reals^{\nx\cdot(\ns\nu+\nx)}$ are introduced such that
\begin{align}
    \uSig\currIter=\Es\zSig\currIter\\
    \xZero\currIter=\Ee\zSig\currIter.
\end{align}
are satisfied.

The optimization problem
\begin{align}
    \minimize{\zSig\nextIter}{J_{j+\ni}}
    \label{Eq: Original Objective Function}\\
    \subjectto & \uSig\nextIter=\Es\zSig\nextIter
    \label{Eq: Define uSig from z}\\
    & \xZero\nextIter=\Ee\zSig\nextIter
    \label{Eq: Define xZero from z}\\
    & \xSig\nextIter=\GMat\uSig\nextIter+\FMat\xZero\nextIter+\dSig
    \label{Eq: Define dynamics}\\
    & \eSup\nextIter=\rSup-\xSup\nextIter
    \label{Eq: Define eSup}\\
    &\uSup\nextIter=\Iu\uSig\nextIter
    \label{Eq: Converged uSig constraint}\\
    & \uSig\currIter=\uSig\nextIter
    \label{Eq: Converged uSig constraint 2}\\
    &\xSup\nextIter=\Ix\xSig\nextIter
    \label{Eq: Converged xSig constraint}\\
    & \xSig\currIter=\xSig\nextIter
    \label{Eq: Converged xSig constraint 2}\\
    & \EfMat\xSig\nextIter=\xZero\nextIter
    \label{Eq: Terminal Constraint}
\end{align}
is now introduced. Here, (\ref{Eq: Define uSig from z}) and (\ref{Eq: Define xZero from z}) define how the input sequence and initial condition are given from the value of $\zSig$. Equation (\ref{Eq: Define dynamics}) defines the lifted state vector for the converged plant and disturbance while (\ref{Eq: Define eSup}) gives the definition of the super-lifted error vector. Equations (\ref{Eq: Converged uSig constraint}) and (\ref{Eq: Converged uSig constraint 2}) constrain the input signal to have converged to a constant value while (\ref{Eq: Converged xSig constraint}) and (\ref{Eq: Converged xSig constraint 2}) constrain the state dynamics to have converged to a constant value. Constraint (\ref{Eq: Terminal Constraint}) requires that the terminal states must be equal to the initial condition from the beginning of the iteration.

To put constraints (\ref{Eq: Converged uSig constraint})-(\ref{Eq: Converged xSig constraint 2}) into better context, note that the cost function is defined over a finite prediction horizon. Consequently, the minimizer of the cost function in the absence of these constraints may be greedy wherein low cost performance is given over the current prediction horizon, but the resulting initial condition at the iteration immediately following the prediction horizon may lead to poor future performance. Hence, the minimizer of the cost function without these constraints may not be desirable over many iterations. By adding these constraints, the solution of the optimization problem will give an input sequence and initial condition such that at convergence, applying the input sequence to the system with the optimal initial condition over a single iteration results in a state trajectory with a terminal condition equal to the initial condition. Thus, this combination of input sequence and initial condition is repeatable such that, at convergence, the same input sequence applied at every iteration will achieve the same performance. Consequently, note that in the optimization problem given by \eqref{Eq: Original Objective Function}-\eqref{Eq: Terminal Constraint}, the constraints imply that the value of the second, third, fifth, and sixth terms on the right hand side of \eqref{eqn:superBlockPerformance} are equal to zero.

The problem given by \eqref{Eq: Original Objective Function}-\eqref{Eq: Terminal Constraint} is an equality constrained optimization problem and can be expressed equivalently as
\begin{align}
    \begin{split}
       \minimize{\zSig\nextIter}{ \frac{1}{2}\zSig\nextIter\QqHat\zSig\nextIter+\qlHat\transpose\zSig\nextIter}\\
    \subjectto &\WMat\zSig\nextIter=\vSig
    \end{split}
    \label{Eq: Compact Optimization Problem}
\end{align}
where $\QqHat,\qlHat,\WMat,\vSig$ are defined in Appendix \ref{apn:EqualityQuadProgMatrices}.

\begin{assumption}
\label{Cond:Positive Definiteness of Compact Quadprog Matrices}
The matrix given by $\QqHat+\WMat\transpose\WMat$ is positive definite.
\end{assumption}
\begin{theorem} Suppose that Condition \ref{Cond:Positive Definiteness of Compact Quadprog Matrices} holds. Then the solution, $\zOpt$, of problem (\ref{Eq: Compact Optimization Problem}) can be found from
\begin{align}
\label{Eq: zopt Definition}
    \begin{bmatrix}
    \zOpt\\
    \lambdaOpt
    \end{bmatrix}=
    \begin{bmatrix}
    \QqHat & \WMat\transpose\\
    \WMat & \textbf{0}
    \end{bmatrix}\inverse
    \begin{bmatrix}
    -\qlHat\\
    \vSig
    \end{bmatrix}.
\end{align}
\end{theorem}
\begin{proof}
As \eqref{Eq: Compact Optimization Problem} is an equality constrained quadratic program, the optimality condition
\begin{align}
    \begin{bmatrix}
    \QqHat & \WMat\transpose\\
    \WMat & \textbf{0}
    \end{bmatrix}
    \begin{bmatrix}
    \zOpt\\
    \lambdaOpt
    \end{bmatrix}=
    \begin{bmatrix}
    -\qlHat\\
    \vSig
    \end{bmatrix}
\end{align}
must hold where $\lambdaOpt$ denotes the vector of Lagrange multipliers. Condition \ref{Cond:Positive Definiteness of Compact Quadprog Matrices} gives that the matrix on the left of the equation is non-singular, thus giving the desired result.
\end{proof}
Therefore, satisfaction of Condition \ref{Cond:Positive Definiteness of Compact Quadprog Matrices} yields a closed form for the optimal value of the input sequence and initial condition at convergence.
\section{Stability for Linear Time-Varying Systems}
\label{sec:LTVStability}

The RHILC control scheme described in Section \ref{sec:RHILC} is developed using a system model given by \eqref{Eq:LTI State Space Model} for application to plants with dynamics given by \eqref{Eq: Plant Dynamics}. Note, however, that the controller could also be developed for LTV system models of the form
\begin{align}
    x^{k+1}\currIter = \AMat^k\currIter x^k\currIter+\BMat^k\currIter u^k\currIter
    \label{Eq:LTV State Space Model}
\end{align}
where $\AMat^k\currIter$ and $\BMat^k\currIter$ denote the time-varying state space matrices. In this case, \eqref{Eq: Gmp Definition} is rewritten as
\begin{equation}
    G^j_{m,p} = 
    \begin{cases}
    \mathbb{0}^{n_x\times n_u}              & m < p         \\
    B_m^j                                   & m = p         \\
    A^{m}_j A^{m-1}_j \hdots A^p_j B^m_j    & \text{otherwise.}
    \end{cases}
\end{equation}
where $j$ denotes the dependency on the iteration. Additionally, \eqref{Eq: Fm Definition} is rewritten as
\begin{equation}
    F^j_m = 
    A_j^m A_j^{m-1} \hdots A_j^2 A_j^1.
\end{equation}
which gives the lifted dynamics
\begin{align}
    \xSig\nextIter=\GMat\nextIter\uSig\nextIter+\FMat\nextIter\EfMat\xSig\currIter.
    \label{Eq: LTV Lifted Dynamics xj}
\end{align}
The super-lifted dynamics
\begin{align}
    \label{Eq: LTV Super Lifted Model}
    \xSup\nextIter=\GSup\currIter\uSup\nextIter+\FSup\currIter\xZero\nextIter
\end{align}
are then given by constructing $\GMat\currIter$ as
\begin{equation}
    \mathbf{G}_j \triangleq \tiny
    \begin{bmatrix} 
\GMat_j & \mathbb{0}            & \hdots    & \hdots        & \mathbb{0} \\
F_j E_F \GMat_j   & \GMat_j         & \hdots    & \hdots        & \mathbb{0} \\
\vdots & \vdots & \vdots & \ddots & \vdots \\
\left(\prod\limits_{m=1}^{N_i-1}F_j E_F\right)\GMat_j & \left(\prod\limits_{m=1}^{N_i-2}F_j E_F\right)\GMat_j & \hdots & \GMat_j & \mathbb{0} \\
\left(\prod\limits_{m=1}^{N_i}F_j E_F\right)\GMat_j & \left(\prod\limits_{m=1}^{N_i-1}F_j E_F\right)\GMat_j & \hdots & F_j E_F \GMat_j & \GMat_j
\end{bmatrix}
\end{equation}
and $\FMat\currIter$ as
\begin{equation}
    \mathbf{F}_j \triangleq \begin{bmatrix}  \FMat\currIter \\ \vdots \\ \left(\prod\limits_{m=1}^{\ni-2}\FMat_j E_F\right)\FMat\currIter \\ \left(\prod\limits_{m=1}^{\ni-1}\FMat_j E_F\right)\FMat\currIter \end{bmatrix}.
\end{equation}
The learning filters are then defined using these updated super-lifted matrices $\GSup\currIter$ and $\FSup\currIter$.

Satisfaction of Condition \ref{Cond: Az Spectral Radius} then gives stability for the system given by \eqref{Eq: zSig Dynamics} with $\Azj$ and $\etaSig\currIter$ given by the updated learning filters.
\section{Illustrative Example: Servo-Positioning System}
\label{sec:SimulationResuls}
The RHILC algorithm is now implemented in simulation to a servo-positioning system described in \cite{Bristow2010}. To demonstrate the advantages of using a receding horizon approach to continuously operating systems, simulations are conducted on an iteration-invariant nominal model using the strategy described in Section \ref{sec:MultiIterationSystemModel}. Additionally, simulations on an iteration varying model with added uncertainty are presented to demonstrate the effectiveness of the algorithm in a more practical case.
\subsection{Simulation of Nominal System}
\label{sec:NominalSimulations}

Implementation of the RHILC algorithm is performed using varying prediction horizon lengths on a nominal model of the servo-positioning system with dynamics given as in \cite{Bristow2010} according to
\begin{align}
\begin{split}
\AMat&=
\setlength\arraycolsep{2pt}
    \begin{bmatrix}
0 &	1\\
-0.71 & 1.50
    \end{bmatrix},\quad
    \BMat=\begin{bmatrix}
    1\\
    1
    \end{bmatrix}
\end{split}.
\label{Eq:Nominal State Space Matrices}
\end{align}
Here, $\ns=50$ and the simulation is run over 10 iterations where each iteration lasts 0.5 seconds. Additionally, the weighting parameters are selected as $q_u=10^{-3},q_{\delta u}=10^{-2},q_e=\begin{bmatrix}1 & 0\end{bmatrix}\transpose,q_x=10^{-6}\cdot\oneVec^{\nx},q_{\delta x}=3\cdot10^{-1}\cdot\oneVec^{\nx},s_x=10^{-18}\cdot\oneVec^{\nx}$ where $\oneVec^a\in\reals^a$ is a vector with all elements equal to 1. Note that in this selection of $q_e$, only the first element is non-zero. Hence, only the tracking error of the first state is penalized. The weighting parameters are chosen such that Conditions \ref{Cond: Az Spectral Radius} and \ref{Cond:Positive Definiteness of Compact Quadprog Matrices} are satisfied.

\begin{figure}[ht]
\includegraphics[width=\columnwidth]{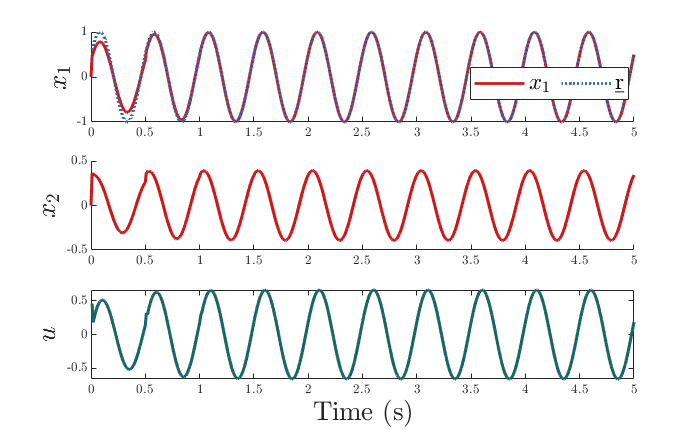}
\caption{State and input sequence history over 10 continuous iterations where each iteration lasts 0.5 seconds with a prediction horizon length of $\ni=3$.}  \label{fig:Nominal Servo State and Input History}
\end{figure}

\begin{figure}[ht]
\includegraphics[width=\columnwidth]{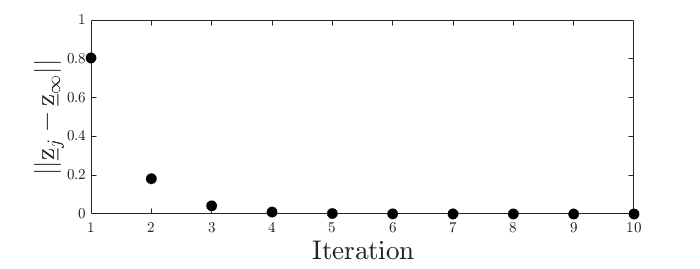}
\caption{Convergence of $\zSig\currIter$ to $\zInf$ for $\ni=3$.}  \label{fig:Nominal Covergence to zinf}
\end{figure}

The state and input histories are shown in Figure \ref{fig:Nominal Servo State and Input History} for the case when $\ni=3$. Here, the input and states appear to have a transient phase before quickly settling to a steady state trajectory. Additionally, state $x_1$ is able to track the reference with small error. In fact, as shown in Figure \ref{fig:Nominal Covergence to zinf}, $\zSig\currIter$ converges monotonically to $\zInf$ as a function of $j$.

\begin{figure}[hb]
\includegraphics[width=\columnwidth]{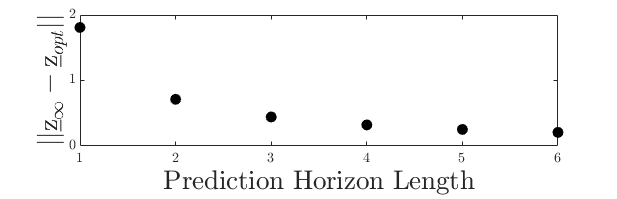}
\caption{The distance between $\zInf$ and $\zOpt$ is reduced as the prediction horizon length increases.}  \label{fig:Nominal Performance vs ni}
\end{figure}

Interestingly, Figure \ref{fig:Nominal Performance vs ni} shows that the measured distances, as defined by the Euclidean norm, between $\zInf$ as given by \eqref{Eq: Definition of zinf} and $\zOpt$ as given in \eqref{Eq: zopt Definition} are reduced monotonically as the length of the prediction horizon is increased. Hence, the RHILC approach allows for improved converged performance in comparison to traditional ILC techniques in which $\ni=1$.


\subsection{Simulation of Uncertain, Iteration-Varying System }
\label{sec:UncertainSimulations}

To demonstrate the effectiveness of the RHILC scheme in a more realistic scenario, the control framework is applied to a simulated servo-positioning system with iteration-varying dynamics and embedded uncertainty.

For these simulations, the iteration-varying state disturbance signal, $\dSig\currIter$, is constructed as Gaussian noise centered at nominal value $\dSig=\IMat_{\ns}\begin{bmatrix} 1.2 & 1.1\end{bmatrix}\transpose$. The plant dynamics are given by 
\begin{align}
    \xSig\currIter = (\GMat^*+\Delta_{\GMat\currIter^*})\uSig\currIter+(\FMat^*+\Delta_{\FMat\currIter^*})\xZero\currIter+\dSig\currIter
\end{align}
where $\GMat^*$ and $\FMat^*$ are given by \eqref{Eq: Gmp Definition} and \eqref{Eq: Fm Definition} with
\begin{align}
    \begin{split}
\AMat&=
\setlength\arraycolsep{2pt}
    \begin{bmatrix}
0 &	1\\
-0.35 & 0.87
    \end{bmatrix},\quad
    \BMat=\begin{bmatrix}
    1.60\\
    0.82
    \end{bmatrix}
\end{split}.
\end{align}
Furthermore, $\Delta_{\GMat\currIter^*}$ and $\Delta_{\FMat\currIter^*}$ represent iteration-varying additive uncertainty that impacts the plant dynamics.

The learning matrices as given by \eqref{Eq: Learning Filter Definitions} are based on the dynamic model
\begin{align}
    \xSig\currIter=(\GMat+\Delta_{\GMat\currIter})\uSig\currIter+(\GMat+\Delta_{\GMat\currIter})\xZero\currIter
\end{align}
where $\GMat$ and $\FMat$ are constructed from \eqref{Eq: Gmp Definition} and \eqref{Eq: Fm Definition} based on the nominal state space matrices in \eqref{Eq:Nominal State Space Matrices} and $\Delta_{\GMat\currIter}$ and $\Delta_{\FMat\currIter}$ are estimates of additive, time-varying changes to the nominal model.

The weighting parameters are updated from Section \ref{sec:NominalSimulations} with $q_{\delta x}=3\cdot10^{-3}\cdot\oneVec^{\nx}$ and $q_{\delta u}=1\cdot10^{-3}$ such that Conditions \ref{Cond: Az Spectral Radius} and \ref{Cond:Positive Definiteness of Compact Quadprog Matrices} are satisfied. The simulation is run using $\ni=\{1,...,6\}$ over 20 iterations.

\begin{figure}[ht]
\includegraphics[width=\columnwidth]{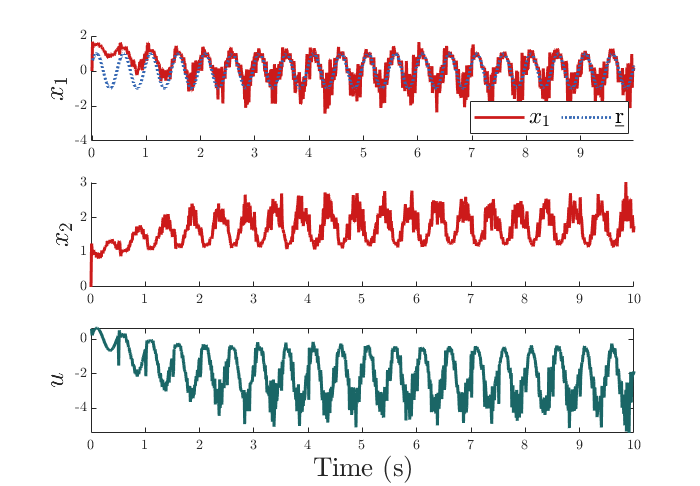}
\caption{State and input sequences over 20 continuously operated iterations of the iteration-varying, uncertain system with a prediction horizon of $\ni=3$.}  \label{fig:Uncertain State and Input History}
\end{figure}

The state and input sequence history is shown in Figure \ref{fig:Uncertain State and Input History} for $\ni=3$.

\begin{figure}[ht]
\includegraphics[width=\columnwidth]{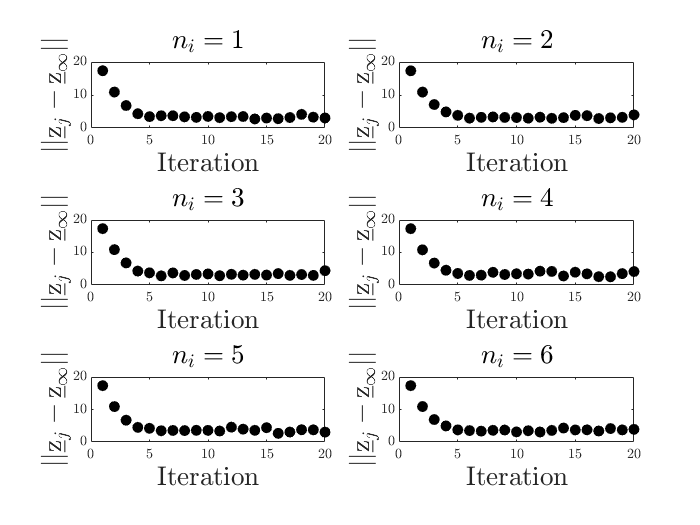}
\caption{Convergence of $\zSig\currIter$ for different values of $\ni$.}  \label{fig:Uncertain Convergence of zinf}
\end{figure}

Additionally, the convergent behavior of the system is examined. Note that in this simulation, $\limit{j}(\dSig\currIter,\Delta_{G\currIter^*},\Delta_{F\currIter^*},\Delta_{G\currIter},\Delta_{F\currIter})=(\dSig,\mathbb{0},\mathbb{0},\mathbb{0},\mathbb{0})$. In Figure \ref{fig:Uncertain Convergence of zinf}, the evolution of $\zSig\currIter$ is depicted for various lengths of the prediction horizon. Note that while these signals tend to converge towards $\zInf$, this evolution is non-monotonic due to the iteration-varying plant uncertainties and disturbances.

\begin{figure}[ht]
\includegraphics[width=\columnwidth]{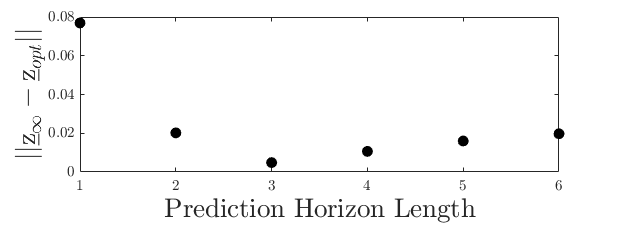}
\caption{The distance between $\zInf$ and $\zOpt$ varies depending on the horizon length.}  \label{fig:Uncertain Performance vs ni}
\end{figure}

A comparison of $\zInf$ to $\zOpt$ as a function of the prediction horizon length is shown in Figure \ref{fig:Uncertain Performance vs ni}. Two key observations are made here. The first is that the use of a multi-iteration prediction horizon outperforms the single-iteration prediction horizon that would be used using standard ILC techniques. Additionally, note that a longer prediction horizon does not necessarily correspond to improved performance at convergence. Hence, an infinite horizon strategy for continuous ILC as proposed in \cite{Lee2001} and \cite{Gupta2006} is not always the best choice when model uncertainties exist.


\section{Conclusions}
\label{sec:Conclusion}
This paper proposes an ILC strategy for continuously operated systems in which the initial condition is not reset between iterations. A multi-iteration dynamic model is defined using a lifted system representation that describes the behavior of the system in response to multi-iteration input sequences and arbitrary initial conditions. A closed form, receding horizon style update law for the input sequence is presented, with stability criteria and desired performance standards established for time-invariant and time-varying systems.

The scheme is then implemented through simulation on a servo-positioning system where it is observed that improved system behavior can be achieved by utilizing a multi-iteration receding horizon approach in comparison to traditional ILC strategies. Additionally, the improved converged performance of the system when utilizing a finite prediction horizon length is established when modelling inaccuracies and disturbances are present. 

Future work includes extensions towards systems with general convex cost functions and for which full state information is not known, and the use of constrained optimal control strategies with terminal components to ensure improved closed-loop performance. Additional work will also address non-linear systems and systems with spatially-defined dynamics.
\section{Appendix}

\subsection{Performance Index Weighting Matrices}\label{apn:perfIndexWeights}
In  \eqref{eqn:longFormPerformance}, the weighting matrices $Q_u$, $Q_{\delta u}$, $Q_e$, $Q_x$, and  $Q_{\delta x}$, are easily defined in terms of three functions,
\begin{itemize}
    \item the ``diagonal'' matrix function, $f_D:\mathbb{R}^{n_a}\mapsto \mathbb{R}^{n_a\times n_a}$ which takes in an input vector, $v=[v_1,\,v_2,\,\hdots ,\, v_{n_a-1},\, v_{n_a} ]$, and outputs a matrix with the entries of the input vector along the diagonal and zeros elsewhere.
    \begin{equation}
        f_D(v) = \begin{bmatrix} 
        v_1 & 0 & \hdots & 0 & 0\\
        0 & v_2 & \hdots & 0 & 0\\
        \vdots & \vdots & \hdots & \vdots & \vdots\\
        0 & 0 & \hdots & v_{n_a-1} & 0\\
        0 & 0 & \hdots & 0 & v_{n_a-1}\\
        \end{bmatrix}
    \end{equation}
    \item the ``repeated block diagonal'' matrix function, $f_{\Delta}:(n_b, \mathbb{R}^{n_{c} \times n_{d}}) \mapsto \mathbb{R}^{n_b n_c  \times n_b n_d}$ which accepts the positive integer, $n_b$, and an arbitrary matrix, $C\in \mathbb{R}^{n_{c} \times n_{d}}$, and returns a block matrix with the input matrix repeated $n_b$ times along the block diagonal
    \begin{equation}
        f_{\Delta}(n_b,C) \triangleq \begin{bmatrix}
        C & \mathbb{0} & \hdots \\
        \mathbb{0} & C & \hdots \\
        \vdots & \vdots & \vdots 
        \end{bmatrix}
    \end{equation}
\end{itemize}

If we then define the vectors of user-selected scalar weights as 
\begin{equation}
\begin{split}
    q_u          & \triangleq \begin{bmatrix} q_{u,1} & \hdots &  q_{u,\nu} \end{bmatrix}^T, \\
    q_{\delta u} & \triangleq \begin{bmatrix} q_{\delta u,1} & \hdots & q_{\delta u,\nu} \end{bmatrix}^T, \\
    q_e          & \triangleq \begin{bmatrix} q_{e,1} & \hdots & q_{e,\nx}\end{bmatrix}^T, \\
    q_x          & \triangleq \begin{bmatrix} q_{x,1} & \hdots & q_{x,\nx} \end{bmatrix}^T, \\
    q_{\delta x} & \triangleq \begin{bmatrix} q_{\delta x,1} & \hdots & q_{\delta x,\nx} \end{bmatrix}^T,\\
    s_{x}        & \triangleq \begin{bmatrix} s_{x,1} & \hdots &  s_{x,\nx} \end{bmatrix}^T.
\end{split}
\end{equation}
Then the gain matrices $Q_u$, $Q_{\delta u}$, $Q_e$, $Q_{\delta e}$, $Q_x$ that encode the relative importance of each term in the performance index are given by
\begin{equation}
    \begin{split}
    Q_u          & = f_{\Delta}\left(n_s,f_D(q_u)\right),\\
    Q_{\delta u} & = f_{\Delta}\left(n_s,f_D(q_{\delta u})\right),\\
    Q_e          & =  f_{\Delta}\left(n_s,f_D(q_e)\right),\\
    Q_x          & = f_{\Delta}\left(n_s,f_D(q_x)\right),\\
    Q_{\delta x} & = f_{\Delta}\left(n_s,f_D(q_{\delta x})\right).
    \end{split}
\end{equation}

\subsection{Cost and Constraint Matrices for Compact Optimization Problem}
\label{apn:EqualityQuadProgMatrices}
In \eqref{Eq: Compact Optimization Problem}, the optimization problem is defined according to
\begin{align}
    \begin{split}
        \QqHat\triangleq&\Es\transpose\left(\Iu\transpose\QuHat\Iu-2\Qdeltau\right)\Es+\left(\GMat\Es+\FMat\Ee\right)\transpose...\\
        &\left(\Ix\transpose(\QxHat+\QeHat)\Ix-2\Qdeltax\right)\left(\GMat\Es+\FMat\Ee\right)\\
        \qlHat\triangleq&(\GMat\Es+\FMat\Ee)\transpose\Big(\Ix\transpose(\QxHat+\QeHat)\Ix\dSig...\\
        &+\Ix\transpose(\sxSup-\QeHat\Ix\rSig)-2\Qdeltax\dSig\Big)\\
       \WMat\triangleq&\Ee-\EfMat(\GMat\Es+\FMat\Ee)\\
       \vSig\triangleq&\EfMat\dSig
    \end{split}.
\end{align}
\section{Acknowledgements}
This work was funded by National Science Foundation grant numbers 1727371 and 1727779 entitled ``Collaborative Research: An Economic Iterative Learning Control Framework with Application to Airborne Wind Energy Harvesting.''

\IEEEtriggeratref{10} 
\bibliographystyle{IEEEtran}
\bibliography{IEEEabrv,main.bib}

\end{document}